\documentclass[aps,prl,onecolumn,showpacs]{revtex4}
\usepackage{graphicx}

\begin{document}

\title{{Comment on Yerin et al , Phys. Rev. Lett. 121, 077002 (2018), and,
Mironov et al, Phys. Rev. Lett. 109, 237002 (2012)}}

\begin{abstract}
In this communication we refute a criticism concerning results of our work %
\cite{BVEprb01} that was presented in references \cite{Buzdin12} and \cite{Buzdin18}.
\end{abstract}

\author{ A.F. Volkov$^{1}$, F.S. Bergeret$^{2}$, and K.B.Efetov$^{1}$}
\affiliation{$^1$Theoretische Physik III, Ruhr-Universit\"{a}t Bochum, D-44780 Bochum,
Germany\\
$^2$Centro de F\'{\i}sica de Materiales (CFM-CSIC), E-20018 San Sebasti\'{a}%
n, Spain}
\maketitle
\date{\today }

%

The authors of references \cite{Buzdin12,Buzdin18} theoretically studied the
possibility of obtaining the Larkin-Ovchinnikov-Fulde-Ferrel (LOFF) state %
\cite{LO65,FF65} in superconductor/ferromagnet (S/F) bi-layers. They
formulated the conditions necessary for realization of such inhomogeneous
state
%

Actually, the idea of inducing the LOFF state in an S/F bilayer is not new
and has been briefly discussed  long time ago, for example,  in Refs. \cite{BVEprl01,BVEprb01}.
 In those publications we have demonstrated a similarity
between an S/F bilayer and a superconductor with an internal exchange field $%
h$. We considered an S/F bilayer with parameters satisfying conditions
\begin{equation}
d_{S}<\xi _{S}\text{, }d_{F}<\xi _{F}  \label{1}
\end{equation}%
where $D_{F}$ and $\xi _{S(F)}$ are  the thickness and correlation length in the  S(F)-layer. 
A highly transparent S/F interface was assumed.
 We have shown that the condensate Green's function in the S film has the same form
as in the LOFF:  $f(\omega )=\tilde{\Delta}/%
\sqrt{(\omega +i\tilde{h})^{2}+\tilde{\Delta}^{2}}$, with renormalized
energy gap $\tilde{\Delta}(T,\tilde{h})$ and exchange energy $\tilde{h}$
[see Eqs.(9-10) in \cite{BVEprl01,BVEprb01} and a discussion there]. This
result demonstrated explicitly that the LOFF state might be realized in such
S/F bi-layers.

The authors of Refs. \cite{Buzdin12,Buzdin18}  extended this idea to the
case of a bilayer with a finite S/F interface transparency and a thickness $%
d_{S}$ that may be comparable or even exceed $\xi _{S}$. They have solved
numerically a simplified effective 1D problem. 

Besides this analysis,  the authors of Refs.\cite{Buzdin12,Buzdin18}
made misleading statements concerning results presented in  our work Ref.\cite%
{BVEprb01}: They write in page 1 of \cite{Buzdin12}: \\
a) ``Recently, this
observation has been questioned in several theoretical works [3--5]
predicting the sign change in the London relation and an unusual
paramagnetic response of the hybrid superconductor or ferromagnet (S=F) and
superconductor or normal metal (S=N) systems''. and in page 2 of \cite%
{Buzdin18}: \\
b) ''It is exactly this FFLO instability which makes impossible
to observe the global paramagnetism predicted in [34-36]. The latter
paramagnetic state just does not correspond to the free energy minimum
[37]''.

We refute both criticisms by  emphasizing that in our work, Ref. \cite{BVEprb01}, we did not calculate
the global response of S/F bilayers to an external magnetic field $H_{ext}$
at all. What we did was to determine the local supercurrents, ${I}_{F}
$, generated spontaneously in the ferromagnet in a S-F bilayer, due to the
magnetic field $\mathbf{B}\cong 4\pi \mathbf{M}_{0}$ {( corrections
to this expression is small in case of a weak proximity effect)}, where $%
\mathbf{M}_{0}$ is the magnetization of the F layer. In other words we 
derived the connection between the Meissner currents in the F layer and the
vector potential
\begin{equation}
\mathbf{I}_{F}=Q_{F}\mathbf{A}_{F}\;.
\end{equation}%
where $\mathbf{A}_{F}=x\mathbf{n\times B}${, } and  $\mathbf{n}${\ is a unit
vector normal to the interface}

We emphasize that the current $\mathbf{I}_{F}$ is not the total current in
S/F bilayer (!), but only the current induced in the F layer. We  found
that indeed the coefficient $Q_{F}$ might change the sign locally due to the spatially 
oscillation of anomalous Green's functions in the F layer. However, in Ref. \cite%
{BVEprb01} we did not made  any  statement about the global response, and therefore the criticism in Ref.\cite{Buzdin12,Buzdin18} is
inappropriate and irrelevant to our results.

\end{document}